\documentclass[aps,prl,twocolumn]{revtex4}
\def\n{{\otimes n}}
\def\m{{\otimes m}}
\def\<{\langle}
\def\>{\rangle}

\def\id{I}

\def\duzomniejsze{<\kern-.7mm<}
\def\duzowieksze{>\kern-.7mm>}

\def\textbf#1{{\bf #1}}
\def\beq{\begin{equation}}
\def\eeq{\end{equation}}
\def\be{\begin{equation}}
\def\ee{\end{equation}}
\def\ben{\begin{eqnarray}}
\def\een{\end{eqnarray}}
\def\beqa{\begin{eqnarray}}
\def\eeqa{\end{eqnarray}}
\def\eea{\end{array}}
\def\bea{\begin{array}}
\newcommand{\bei}{\begin{itemize}}
\newcommand{\eei}{\end{itemize}}

\def\dcal{{\cal D}}
\def\bcal{{\cal B}}

\def\trace{\mbox{Tr}}
\def\dt#1{{{\kern -.0mm\rm d}}#1\,}
\def\ypodpis{\raise4mm\hbox{$\omega$}}

\begin{document}
\title{Local information as a resource in distributed quantum systems}
\author{Micha\l{} Horodecki$^{(1)}$, Karol Horodecki$^{(2)}$, Pawe\l{}
Horodecki$^{(3)}$, Ryszard Horodecki$^{(1)}$ 
Jonathan Oppenheim$^{(1)(4)}$, Aditi Sen(De)$^{(1)}$ and 
Ujjwal Sen$^{(1)}$}

\affiliation{$^{(1)}$Institute of Theoretical Physics and Astrophysics,
University of Gda\'nsk, Poland}
\affiliation{$^{(2)}$Faculty of Mathematics, University of Gda\'nsk, Poland}
\affiliation{$^{(3)}$Faculty of Applied Physics and Mathematics,
Technical University of Gda\'nsk, 80--952 Gda\'nsk, Poland}
\affiliation{$^{(4)}$
Racah Institute of Theoretical Physics, 
Hebrew University of Jerusalem, Givat Ram, Jerusalem 91904, Israel}

\begin{abstract}
A new paradigm for distributed quantum systems where {\it information}
is a valuable resource is developed. 
After finding a unique measure for information,
we construct a scheme  for it's manipulation
in analogy with entanglement theory. In this scheme
instead of  maximally entangled states, two parties distill {\it local}
states. We show that, surprisingly, the main tools of entanglement
theory are general enough to work in this opposite scheme.
Up to plausible assumptions, we show that the amount of information
that must be lost during the protocol of concentration of local information
can be expressed as the relative entropy distance from 
some special set of states.

\end{abstract}
\maketitle

The notion of quantum
correlations is a more general than 
entanglement\cite{Bennett-nlwe,discord}. 
A formal measure of quantum correlations in measurements
(quantum discord) \cite{discord}
was found, based primarily on an entropy-like function.
Recently, the first {\it operational} approach
to quantify quantum correlations
was introduced in \cite{OHHH2001}.  Subsequently, a 
similar approach was used to justify a physical interpretation of 
(the optimized) quantum discord \cite{Zurek-demons-02}.
The results of \cite{OHHH2001}
was based on the idea that using  a system in a pure state one can 
draw work
from a single heat bath. This scenario was used in the case of distributed 
quantum systems: Alice and Bob share a state, have local heat baths,
and can use only local operations and classical communication (LOCC), 
to concentrate the information contained in the
state, in order to draw work. The amount of work drawn by LOCC
is usually smaller than the one extractable if Alice and Bob 
can use global operations. The resulting difference denoted
by the deficit $\Delta$ accounts for the part of correlations that must be lost
during classical communication, thus describing purely quantum correlations.
In the case of $\Delta$ for pure states, it was argued to be
exactly equal to the entanglement while for mixed states it is
supposed to be an independent quantity. In this context, it is clear
that understanding the problem of concentration of information 
will provide valuable insight into the nature of quantum correlations.
Yet the early development of these ideas \cite{OHHH2001,compl} 
indicated that the proposed scenario is completely different 
that anything we had in quantum information theory so far.
In particular the serious difficulty (which shall be removed in
this paper) was that one is not even able to obtain (without
additional assumptions) the value of $\Delta$ in the simplest case
of a two-qubit Bell state.

In this context basic questions arise: (i) What is the connection
between the above thermodynamical quantification of quantum correlations
and the main concepts of quantum information theory?
(ii) 
Can we formulate the concentration of information within a framework of manipulating resources
like in entanglement theory? An even more basic question is: (iii) can we 
give up thermodynamics, and formulate the problem solely
in terms of quantum information?
To see the importance of the latter, let us note that the discovery
of teleportation and Shor's algorithm was possible
by restricting to the naked quantum mechanical formalism.

A powerful domain of quantum information
theory which is  well 
formalized is entanglement theory where the primitive
 notions are only states 
of compound systems and a class of operations.
One would like to formalize
our scheme in a similar manner
An important question is then:
can we make our scheme similar to entanglement theory? 
Apart from conceptual benefits,
one then hopes that the powerful tools of entanglement theory can be 
borrowed to tackle information manipulations.

These questions, answered in this work,  
are of fundamental importance for investigations
of quantum correlations of composite states. 
First, we will not start from an a priori notion
of information, and will not assume that we are interested
in the usual function for information $I=N-S$  ($N$ is the number of qubits
associated with a given state \cite{N} and $S$ is its von Neumann entropy)
which we used in \cite{OHHH2001} based on
thermodynamical considerations. Instead we define a class of global operations
over a quantum system  called Noisy Operations (NO), and define information
as whatever quantity does not increase under the class.  We find that, under certain
natural assumptions $N-S$ is the {\it unique measure of information}.

Then we pass to distributed quantum systems. In this context, a natural class of operations
is LOCC\cite{BDSW1996}.
To tackle information, we restrict this class, obtaining N(oisy)LOCC.
It differs from  LOCC in that only maximally mixed local ancillas
can be added for free.
We show that the task of concentration of information is a dual scheme 
to the entanglement distillation scheme. Instead of maximally entangled states,
Alice and Bob have to draw {\it local states}. In entanglement theory, pure 
local states were a free resource; in our scheme, they represent 
a useful resource: {\it localized information}.  
The schemes are then quite opposite. Surprisingly, however, we show that the  
main tools of entanglement theory can be applied to the problem of 
concentration of information. This is due to the fact that both paradigms
are based on the general scheme  of state transformations under a given class of 
operations.  Using  the technique of monotones
 \cite{Vidal-mon2000,limits} and  Rains semidefinite program \cite{Rains2001a}
we obtain rigorously that for pure states, the amount of localisable information $I_l$
is equal to $N-E$, where $E$ is the entanglement. This means that the
quantum correlations defined by the deficit $\Delta=(N-S) - I_l$ for pure states
are equal to its entanglement.  Also up to plausible assumptions we obtain, that $\Delta$
is the relative entropy distance from some special set of states (smaller than the 
set of separable states; see \cite{Vedral1999} in this context).

{\it Unique measure of information:}
Let us begin by looking
at the class of allowable operations for a single system.
We use {\it noisy operations} (NO)\cite{Karol}.
We imagine that the Universe
is mostly filled with qubits in maximally mixed states.
This is a reminiscent of the heat bath in thermodynamics.
Then a qubit in a
non-maximally mixed  state is valuable. Such qubits cannot
be brought in for free, but {\it only manipulated}.
Accordingly we allow a) unitary operations, b) adding an ancilla in
the maximally mixed state and b) rejecting (tracing out) part
of the system. Operations a) and b) are reversible while c) is not.
Note that we do not consider measurement as a separate operation, as the
latter involves implicitly a device with memory initially in a pure state.
Rather we will count the device qubits explicitly i.e. treat it as part
of the system. The measurement is then a unitary transformation.
We can now ask, under NO, what is the number of pure states which
can be distilled from a state $\varrho$.
It turns out that the number of pure states
which can be distilled from a N-qubit state $\varrho$ under NO is
the information $I(\varrho)=N-S(\varrho)$ where
$S(\varrho)=-\trace \varrho\log\varrho$ is the von Neumann entropy
and that this process is reversible. 
The details of the proof of this will be presented in \cite{uniqueinfo},
so here, we just sketch an outline.
The method of distilling pure states is essentially
Schumacher compression \cite{Schumacher1995}
(cf. \cite{PetzMosonyi} and references therein).
In the original compression scheme, the pure qubits were rejected,
while the qubits carrying the compressed state were kept, as the latter was the signal.
Here, we do the {\it converse}: keep the pure part (the mixed state
is not interpreted as signal, but as noise, as in \cite{VaziraniS1998}).
Therefore it is cooling rather than compression \cite{Beth-thermo}.

One can also show, that the converse protocol is possible: i.e.
to create $n$ copies of state $\varrho$ one needs $n(N-S(\varrho))$
pure qubits and $nS(\varrho)$ qubits of noise.
(This is somewhat similar to the much more advanced reverse
second Shannon theorem problem \cite{Shor}).

It follows that in the asymptotic regime (many copies), apart from
the maximally mixed state
any state can be reversibly converted into any other, at a rate
fixed by the entropy of the state. Then, as with the Carnot efficiency
(c.f. \cite{popescu-rohrlich})
there is a unique function that is monotonic under transitions
possible by NO (up to a factor and an additive constant).
The function we can call  {\it information}, as everybody will
agree that information, whatever it is, cannot increase under NO.
We can fix the free parameters, by requiring that for pure N-qubit states
$I=N$. {\it We have thus obtained that I is the unique measure of information
contained in quantum systems:}
Information cannot be created. It can only be manipulated or lost.

{\it Concentration of information to a local form:}
Now we are in a position to pass to distributed systems shared
between Alice and Bob (for simplicity we
will deal with bipartite systems).  We will now restrict the class of
operations (called NLOCC) allowing (i) local noisy operations
(ii) sending qubits down dephasing channel
which is equivalent to classical communication.
The only difference between LOCC and the present
NLOCC class is that in the former, local ancillas
are free while in the latter, only maximally mixed
ancilla are free \cite{clocc}.

In entanglement theory
the valuable resource was the maximally entangled state $\psi_+^d={1 \over \sqrt{d}}
\sum_{i=1}^d|i\>|i\>$
and one asked: {\it How many maximally entangled states
can be distilled from $\varrho$ with help from LOCC operations?}
Instead, we propose to care
about {\it local information}. Consider a bipartite system with
information contents $N-S(\varrho)$. The main question
is \cite{OHHH2001}: {\it How much of the information can be concentrated
to local form by NLOCC ?}.
Since we know that local information is equivalent to
local pure states,
the question can be converted into: {\it How many
pure product states can one distill from the state $\varrho$
under the NLOCC class?} From the formal
point of view this happens to be very similar to the
problem of entanglement distillation.
The key difference is that instead of
maximally entangled state, we want to distill states that
were regarded as useless so far! (The character of the distilled 
resource naturally imposes the use NLOCC instead of LOCC.)
Surprisingly, this allows us to quantify quantum correlations:
the amount of information that cannot be concentrated must be
quantum, because it is destroyed by the dephasing channel.
Thus we define the classical (i.e. local) information contents  $I_l$
of the state $\varrho$, as the optimal rate $2m/n$, where $n$
is the number of copies in the state $\varrho$,   $m$ is the number of two-qubit
pairs output into pure product states (we put the factor 2 since we 
count local information in bits).
The contents of quantum correlations $\Delta$ is then defined as
the difference: $\Delta=I - I_l$.
The task of evaluating $I_l$ (hence also $\Delta$) turned
out to be surprisingly hard. Even for the state $\psi_+^d$,
we were not able to prove rigorously, that
$I_l=\log d$ \cite{OHHH2001} which is intuitively obvious.

Below, we find that since $I_l$ is some conversion rate under NLOCC,
the following powerful tools of entanglement theory (with
suitable modifications) can be applied: enlarging the class of operations,
Rains approach \cite{Rains2001a}  and the concept
of monotones  \cite{Vidal-mon2000,limits}.

First of all,  NLOCC operations (like LOCC ones)
are hard to deal with from a mathematical point of view.
In entanglement theory, one considers the greater classes of
operations\cite{Rains2001a,PlenioVedral1998}:
separable operations and so-called PPT operations\cite{Rains2001a}.
To obtain analogous classes, we only need to add a condition of
{\it preserving maximally mixed state} (PMM). Thus we can optimize the rate
over PMM versions of the above classes, which will be a
great simplification, and will lead to useful bounds
for $I_l$.

In the Rains approach, one assumes a fixed rate of conversion
from one state to another, and evaluates the fidelity
of the conversion. The rate is attainable if the
fidelity can approach $1$ in the limit of many input copies. For simplicity
we will work with two-qubit states. We fix the rate $r$, which means that
for $n$ input copies we will obtain $m= nr/2$ output pairs. The pairs
are in a final joint state $\varrho'=\Lambda(\varrho^\n)$, where
$\Lambda$ is the optimal NLOCC operation. 
We want to estimate the fidelity
between $\varrho'$ and the $m$ copies of the two-qubit
state we would like to have (e.g. $P_{00}=|00\>\<00|$)
for fixed $n$
\be
F=\trace[ P_{00}^\m \Lambda(\varrho^\n)].
\label{fidelity1}
\ee
To evaluate $F$ one can use techniques developed in
\cite{Rains2001a}.
One rewrites (\ref{fidelity1}) in the form
$F=\trace[ \Lambda^\dagger (P_{00}^\m )\varrho^\n]\equiv \trace\Pi
\varrho^\n$, where $\Lambda^\dagger$ is the dual map \cite{dual}.

Let us find constraints for the operator $\Pi$. One can
check that since $\Lambda$ is trace preserving
then $0\leq \Pi\leq \id$. Since $\Lambda$ preserves
maximally mixed state (ie. $\Lambda(I/d_{in})=I/d_{out}$),
one finds that ${d_{in}\over d_{out}}\Lambda^\dagger$ is trace preserving;
in our case $d_{in}=2^{2n}, d_{out}=2^{2m}$,
hence $\trace\Pi= 2^{n(2-r)}\equiv K$, so that $\Pi/K$ is a state.
Since $\Lambda$ is NLOCC, so is
$\Lambda^\dagger$. Then, since $P_{00}^\m$ is separable, so is $\Pi/K$.
This constrained optimization problem can be treated
using the duality concept in semidefinite programming as
in \cite{Rains2001a}. We will do so elsewhere. Here
we prove the following bound:

{\it Proposition 1. - For any bipartite state $\varrho_{AB}$ of
$N=N_A+N_B$ qubits, we have
\be
I_l\leq N- S_\infty(\varrho_X), \quad X=A,B
\ee
(or, equivalently, $\Delta \geq S(\varrho)-S_\infty(\varrho_X)$)
where
$S_\infty= -\log \lambda_{\max}$, with $\lambda_{\max}$ being
the largest eigenvalue of $\varrho_X$.}

{\it Remark - } Since usually $S_\infty <S$, this bound
is weaker than the one proven under an assumption in \cite{OHHH2001}.
However for the maximally entangled state it is tight.

{\it Proof - } We prove for two-qubit states (generalization
to higher dimensions is obvious). We note that ${1\over K}\Pi$ is a separable state. Then
$F\leq K\sup_\sigma \trace \sigma \varrho^\n$ where sup is taken over
separable $\sigma$, which can be taken pure. Consequently
$F\leq K \sup_{\psi\otimes \phi}
\<\psi \otimes \phi| \varrho^\n |\psi \otimes \phi\>\leq K\sup_\psi
\<\psi|\varrho_X^\n|\psi\>=K\lambda_{\max}^n$. Since we require $F\to1$ \
for large $n$, we must have $r\leq 2 + \log\lambda_{\max}$ which gives
the expected bound.

In \cite{OHHH2001} we argued that
$I_l(\psi)=N-S_X(\psi)$ which is equivalent to
$\Delta(\psi) =S_X(\psi)$ where $S_X(\psi)$ is either
of the reductions of $\psi$ and is equal to its
entanglement $E(\psi)$. Here we will prove
it rigorously. Following Prop. 1 one has $I_l(\psi_+^d)\leq \log d$. This
bound is attainable\cite{OHHH2001}. Since $N(\psi_+^d)=\log d^{2}$
the rigorous proof for maximally entangled states is complete.
To extend it to other pure states, we use a version of
the entanglement dilution scheme\cite{BBPS1996}, in which out of
$nS_A(\psi)$ shared pairs $\psi^{2}_{+}$
one gets $n$ pairs in state $\psi$ by use
of $O(\sqrt n)$ bits of communication\cite{LoPopescu}.
It follows that due to asymptotically negligible communication cost,
production of entropy is also negligible.
Thus, to create $\psi^\n$ one needs $m=nS_A$
copies of $\psi^{2}_{+}$ (which occupy $2m$ qubits) plus $Nn-2m$ pure
local qubits. Now if from $\psi$ one could draw more than $N-S_A$
per pair, then one could draw more local bits than $m$ from
the maximally entangled state, converting them first to $\psi$ and then drawing
local bits from $\psi$. On the other hand, there is an obvious protocol
to get $N-S_A$ local bits: Alice, sends her half to Bob,
via the dephasing channel \cite{OHHH2001}.
Thus $I_{l}(\psi)=N-S_{A}(\psi)$ or, equivalently,
$\Delta(\psi)=S_{A}(\psi)$ which concludes the proof,
since $S_{A}(\psi)=S_{B}(\psi)$.
For the tripartite GHZ state, one can likewise
show that $\Delta=1$.

We will now proceed to estimate $I_l$ from below. To this end we need
the notion of {\it Implementable Product Basis} (IPB). It is a basis that
can be achieved from the standard basis $|i\>_A|j\>_B$ by means of reversible
NLOCC operations. (Of course, since we want to transfer pure states
into pure ones, local noise is here not needed.)
IPB is always distinguishable by LOCC. Hence the basis given in
\cite{Bennett-nlwe} is not IPB. Most likely the converse is also true.
It is obvious
that a state with an IPB eigenbasis (call it an IPB state)
has $\Delta=0$. There is also a natural scheme of information
concentration for other states: one fixes some IPB (let it
be $\bcal$), and dephases a given state $\varrho$ in this basis
which, by definition, is possible by NLOCC. The final state
$\varrho'$ has $\Delta=0$, so one can draw
$I_l(\varrho')=N-S(\varrho')$ local pure qubits. However
$S(\varrho')=H(\varrho,\bcal)$ (Shannon entropy of
$\varrho$ in basis $\bcal$). Thus we have the bound
$I_l(\varrho)\geq N - \inf_{\bcal\in IPB} H(\varrho,\bcal)$.

Since one has  $\inf_{\bcal\in IPB} H(\varrho,\bcal)=S(\varrho) +
\inf S(\varrho|\sigma)$ \cite{wyprowadzenie}, where infimum is taken over IPB states $\sigma$,
and $S(\varrho|\sigma)=\trace\varrho\log\varrho - \trace \varrho \log \sigma$,
and since it may be better to operate collectively on $\varrho^\n$ one
obtains

{\it Proposition 2. - For $N$ qubit bipartite state $\varrho$ one has
\be
I_l \geq N - S(\varrho) -\dcal^\infty(\varrho)
\ee
(or, equivalently, $\Delta \leq \dcal^\infty(\varrho)$)
where $\dcal^\infty(\varrho)$ is regularized relative entropy
distance from the set of IPB states $\dcal=\inf_{\sigma\in IPB}S(\varrho|\sigma)$
\cite{reg}.}

Now we make an attempt to show equality by using
the method of monotones. As shown in \cite{DonaldHR2001,Michal2001}
(see also \cite{limits})
if a function $M(\varrho)$ is a) non-increasing under a
class of operations b) asymptotically continuous,
then its regularization satisfies
\be
M^\infty(\sigma) R(\varrho\to\sigma) \leq M^\infty(\varrho)
\label{mono-rate}
\ee
where $R$ is the optimal transition rate.
As $M$ we will take $N - S(\varrho) -\dcal(\varrho)$.
$S(\varrho)$ satisfies b) by Fannes inequality \cite{Fannes1973}, while
by \cite{DonaldH1999} also $\dcal$ satisfies b) \cite{cont-of-D}.
In our case $\sigma$ is the two-qubit product state $\sigma=P_{00}$
so that $M(\sigma)=M^\infty(\sigma)=2$. Since we intend to count not
product states, but bits, we have $I_l=2R$. The inequality then
reads
$I_l(\varrho) \leq M^\infty(\varrho)$.
Now if $M$ is monotonic (satisfies a)), then combining 
Prop. 2 with the inequality we would know $I_l$
{\it exactly}. Let us then check monotonicity. To this end 
we use the formula $M(\varrho)= N - \inf_{\bcal\in IPB}H(\varrho,\bcal)$
and consider sending down the dephasing channel
in two stages: (i) local dephasing of part of the system (ii)
sending dephased part via an ideal quantum channel. It is easy to see that
operations (ii) as well as local unitary transformations
do not change $M$ (they change one IPB into other one).
From the fact that a product of an IPB with some local basis
is again an IPB it follows that partial trace does not increase $M$.
We are not able to prove monotonicity of $M$ under
adding noise
(ie. maximally mixed states) and local dephasing.
Concerning noise, it is
rather unlikely that it can increase $M$, and moreover,
most likely it is not needed in concentration task at all
(basically noise is needed when we want to create more mixed states
out of less mixed ones).
However the question of whether dephasing can increase $M$ remains elusive.
This is very closely related to the assumption we made in \cite{OHHH2001}
.There we would obtain some improvement
of the bound of Prop. 1. Here if the assumption was true we would obtain
the exact formula for $\Delta$. The problem can be formulated as follows:
can Alice in the optimal information concentrating protocol,
partially dephase her system, and at some later stage, dephase
again, so that two dephasings do not commute? 
Since non-commuting measurements destroy
information, we believe the answer
is ``no''. This justifies the following conjecture:

{\it Conjecture .- For $N$ qubit bipartite state, the amount
of concentratable information is
\be
I_l= N -S(\varrho) - \dcal^\infty(\varrho).
\ee}
Thus we would have that $\Delta=\dcal^\infty(\varrho)$ ,
i.e. $\Delta$ is the (regularized) relative entropy distance form
IPB states. The information $I_l$ is then given in terms of
a Shannon entropy, an idea which arose naturally in \cite{compl}.

In conclusion we have derived the notion of information
form general principles. Then we have shown that local information
$I_l$ can serve as a resource in distributed systems in an analogous
way to entanglement theory. We were then able
to use powerful tools developed in the latter, with suitable modifications.
We obtained bounds for both the amount of information
that can be concentrated to local form  ($I_l$) and
the corresponding deficit $\Delta$. Under some assumption we
argued that the deficit $\Delta$ is the relative entropy distance
from the set of states having $\Delta=0$.
Since the deficit is a measure of quantum correlations
based on thermodynamics, the present program
can be viewed as a bridge that links quantum
correlation theory and quantum thermodynamics
in a systematic way. In particular there is a hope 
to explain nonlocality without entanglement
in a thermodynamical manner,
as nontrivial  states diagonal in basis of Ref. 
\cite{Bennett-nlwe} may have $\Delta >0$.   


{\bf Acknowledgments}:
This work is supported by EU grant EQUIP, No. IST-1999-11053.
J.O. is supported by
grant No. 129/00-1
of the Israel Science Foundation.


\begin{thebibliography}{31}
\bibitem[{dua()}]{dual}
\bibinfo{note}{The dual map to a map $\Lambda$ is defined as unique map
  $\Lambda^\dagger$ satisfying $\trace A^\dagger \Lambda(B)=\trace
  \Lambda^\dagger(A)^\dagger B$. For completely positive map
  $\Lambda(\cdot)=\sum_i V_i(\cdot)V_i^\dagger$ we have $\Lambda^\dagger
  (\cdot)= \sum_i V_i^\dagger(\cdot)V_i$.}
\bibitem[{con()}]{cont-of-D}
\bibinfo{note}{The set of IPB states is not convex as assumed in Ref.
  \cite{DonaldH1999}. However it is enough that the set is star-like, so that
  any IPB state can be connected with maximally mixed state by a straight line
  that is inside the set.}
\end{thebibliography}

\end{document}